# Role of temperature oscillation in growth of large-grain CdZnTe single crystal by traveling heater method

P. Vijayakumar[a], Subham Dhyani[a], K. Ganesan[b,c], R. Ramar[a,c], Edward Prabu Amaladass[b,c], R.M.Sarguna[a], S. Ganesamoorthy[a,c,*]

[a] *Safety, Quality & Resource Management Group, Indira Gandhi Centre for Atomic Research, Kalpakkam-603102, India*

[b] *Materials Science Group, Indira Gandhi Centre for Atomic Research, Kalpakkam-603102, India*

[c] *Homi Bhabha National Institute, Training School Complex, Anushakti Nagar, Mumbai, 400094, India*

[*]Corresponding author. Email: sgm@igcar.gov.in

**Abstract**

Self-nucleation in CdZnTe crystal growth remains a significant challenge, despite numerous attempts to achieve large-grain single crystals by restricting multi-nucleation during growth process using the traveling heater method. In this study, we present a novel approach to achieve large-grain CdZnTe single crystals by introducing temperature oscillations above the crystallization temperature during the growth process. This method effectively suppresses secondary nucleation and promotes the preferential selection of a single grain during early stage of growth as well as along the growth axis, by reducing multi-nucleation. By adjusting the amplitude and the number of temperature oscillations, we have successfully grown CdZnTe single crystals with dimensions of 20 mm in diameter and 60 mm in length. The resulting crystals exhibited excellent compositional homogeneity, with a nearly constant resistivity of ~$10^9$ Ω·cm and Te inclusions smaller than 15 μm along the growth axis. Additionally, the crystal elements were of detector grade achieving an energy resolution of 4.5% for gamma radiation at 662 keV from a $^{137}$Cs source in a quasi-hemispherical geometry. This study highlights the critical role of temperature oscillations in controlling secondary nucleation and promoting the formation of large-grain single crystals.

## 1. Introduction

$Cd_{0.90}Zn_{0.10}Te$ (CdZnTe) based gamma radiation detector has attracted a lot of attention in recent decades due to its potential use in nuclear radiation detection, because of the improved energy resolution at room temperature operation compared to HPGe detectors [1]. Gamma spectrometers using CdZnTe are excellent candidates for various industrial applications, including medical imaging, the nuclear spectroscopy, security scanning and astronomy [2–4]. The sensitivity and the resolution of the gamma radiation detectors can be improved by using CdZnTe single crystal elements with good crystallinity, high resistivity, and a low concentration of Te inclusions. However, the growth of large-grain high-quality single crystals involves several technological challenges due to the inherent properties of the raw material like (i) extremely low thermal conductivity resulting in a high segregation co-efficient for Zn; and (ii) high vapour pressure of Cd resulting in defect formation and growth instabilities at the solid-liquid interface. These factors lead to the formation of Te inclusions, compositional inhomogeneity and other structural defects which are detrimental to the performance of the device. Although conventional melt growth techniques, such as Bridgman, high-pressure Bridgman, and vertical gradient freeze [5,6], are capable of producing large CdZnTe crystals, they fail to address several critical challenges. These limitations can be effectively mitigated by employing low-temperature solution growth using the traveling heater method (THM), which involves melting only a small portion of the charge via a solvent. Overall, comparative assessments indicate that THM-based growth provides significant advantages over traditional melt growth methods.

The thermal conductivity of molten and solid phase CdZnTe is 0.06 and 0.03 W/m.K, respectively [7,8]. The low thermal conductivity of crystalline CdZnTe hinders the efficient removal of latent heat during crystal growth, leading to multi-nucleation. Additionally, the high segregation coefficient of Zn in the CdZnTe matrix ($K_{eff}$ ~1.35) induces compositional inhomogeneity that leads to the formation of multi-crystalline ingots [9]. THM is a well-established technique for growing detector-grade CdZnTe crystals since it enables precise control over crystal growth parameters at a relatively low growth temperature. Moreover, THM yields high quality CdZnTe single crystals with homogeneous composition and low defect density. The self-nucleation method is a simple and widely used method to grow single crystals since it eliminates the demand for seed crystal. In the conventional self-nucleated THM, the growth begins with multi-nuclei formation at the conical tapered region and then, the growth dominates along the preferential crystal axis that eventually grow as large grain CdZnTe crystal. However, some reports suggest the formation of secondary nucleation during growth that results in multigrain structure which reduces the usable volume of CdZnTe crystal. This problem can be rectified by employing a seed crystal at the bottom of the crucible. However, there exists additional difficulties in employing seed crystal in THM system. Furthermore, even with the use of a seed crystal, achieving single-grain growth throughout the crystal volume is not always guaranteed, primarily due to the poor thermal conductivity of the CdZnTe crystal [7,8]. The promotion of single nucleation and the suppression of multi-nucleation in self-nucleated THM growth continue to pose a significant challenge for researchers [10]. The growth of large grain CdZnTe single crystals with improved structural quality was attempted by various kinds of modifications, such as introduction of bell-curved temperature profile of the heater which makes convex solid-liquid interface [11], modified Bridgman method by super heating the melt before growth [12], temperature oscillation around crystallization point and external cooling support [13,14], which help in minimizing the multi-grain formation during the initial stage of growth. Moreover, it is also essential to optimize growth parameters such as temperature gradient, growth rate, and melt composition in order to achieve high quality single crystal.

The temperature oscillation method (TOM) is one of the oldest techniques utilized to enhance size of the single crystals [15–20]. Here, the temperature at the growth interface is periodically oscillated around the crystallization point. This temperature oscillation leads to periodic re-melting and re-growth that can improve crystalline quality by reducing the number of grains, encouraging the formation of larger single grains, and enhancing uniformity throughout the crystal. Besides, such a temperature variation can also occur spontaneously at the growth interface as a result of melt convection during crystal growth [21]. However, in the TOM technique, the temperature of the melt near the solid-liquid interface is deliberately oscillated over a mean value. TOM was successfully utilized to increase the grain size of metal halide single crystals from vapour in 1970s [15]. Also, a five-fold increase in size of certain dichalcopyrite single crystals were achieved by iodine vapour transport method through periodic temperature oscillation at the growth interface [16]. Mroczkowski et al. [22] had reported that back-melting, followed by controlled resolidification, reduces dislocation density in GaSb crystals. Also, Van Run [17,18] had examined the impurity striation pattern associated with periodic remelt and also, the critical pulling rate to suppress such periodic remelt. Crochet et al. [19] analyzed thermomechanical aspects of GaAs melts and predicted the periodic freezing and remelting at the interface in presence of temperature oscillation using numerical simulation. Masalove et al. [20] studied the role of hydrodynamics and oscillation of temperature in oxide single crystal growth from high temperature solutions with use of accelerated crucible rotation technique. Recently, Attolini et al. [23] has reported an increase in Ge crystal size and reduction in number of crystalline nuclei during the primary nucleation using oscillatory temperature profile under chemical vapour transport technique. Overall, the occurrence of temperature oscillation near the growth interface, either induced intentionally or developed naturally, has profound impact on the quality and size of the crystals. Especially, the intentionally induced temperature oscillation is quite popular and effectively utilized to increase the size of the crystals under vapor transport method whilst it is less popular in the case of melt or high temperature solution growth.

In this study, we deliberately introduced temperature oscillations in THM method by raising the furnace temperature above the growth temperature ($T_G$) with an oscillation amplitude ($\Delta T$) ranging from 3 to 15 °C. Upon reaching the $\Delta T$, the furnace temperature was reduced to the growth temperature and then, maintained. The oscillation amplitude and the number of oscillations were optimized to enhance the likelihood of forming large-grain CdZnTe single crystals using the THM. Using the optimized parameters for temperature oscillation, multi-nuclei formation has been minimized resulting in the growth of a sizeable, high-quality, detector-grade CdZnTe single crystal. The crystal grown with optimized parameters were then analyzed using various analytical techniques and evaluated for gamma-ray spectroscopic applications.

## 2. Experimental
### 2.1. Crystal growth

Growth of CdZnTe single crystal was carried out using in-house developed THM system. Initially the growth parameters such as solute-solvent ratio, growth temperature, temperature gradient at the solid-liquid interface, ampoule rotation rate and growth rate were optimized using the quartz ampoule with inner diameter of 20 mm and the details are discussed elsewhere [24]. One limitation of this method is that it does not always lead to the selection of a single grain. In this study, we explored the influence of temperature oscillation on the growth of large grain CdZnTe crystal by employing different oscillation amplitudes and the number of oscillations. Towards this goal, we conducted four independent growth runs, named as Run-1, Run-2, Run-3, and Run-4, using different temperature oscillation amplitude whose profile is given in Fig. 1a. Prior to applying temperature oscillation, we selected optimized growth parameters including growth rate of ~ 4 mm/day and temperature gradient of ~ 45 K/cm, as detailed in Ref [24]. The saturated solution, consisting of Te solvent and CdZnTe solute, was maintained at the growth temperature of ~ 850 °C. Subsequently, the growth was initiated by translating the heater assembly, following one of the temperature profiles shown in Fig. 1a. Moreover, a controlled amount of indium dopant was incorporated into the initial charge in all CZT growth runs to effectively compensate for inherent cadmium vacancies.

In Run-1, no temperature oscillation was applied, similar to conventional THM growth. For Run-2 and Run-3, temperature oscillations were applied with an amplitude, $\Delta T$, of 3 and 10 $^0$C, respectively, for 5 and 10 cycles at the beginning of the growth. In Run-4, the temperature oscillations were performed for 40 cycles with a $\Delta T$ of 15 $^0$C from the growth temperature. The furnace temperature was raised at the rate of ~ 2, 20 and 15 $^0$C/h and subsequently after reaching $\Delta T$, it was reduced to actual growth temperature at the rate of ~ 0.7, 1.1 and 1.4 $^0$C/h for the growth runs Run-2, Run-3 and Run-4 respectively, as can be seen from Fig. 1b which displays the actual temperature oscillation profile as a function of time. The time period for each cycle was kept constant as 24 hours for the growth runs 2, 3 and 4. Further, Table 1 provides the details of the temperature oscillation and dwell time used for these growths. During the upslope in the temperature oscillation process, the initially nucleated and grown grains undergo partial re-melt due to the under-saturated Te+CdZnTe solution.

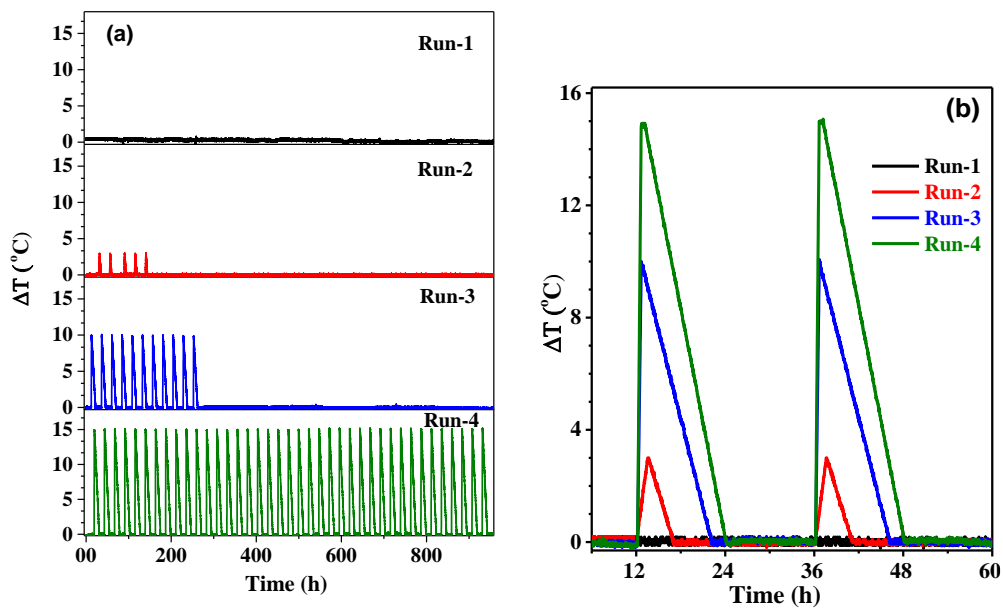

Fig.1. (a) Temperature oscillation profiles, and (b) A closer look at rise and fall of temperature with time for four different growth conditions

Conversely, during the partial downslope near the growth temperature and during the dwell time, supersaturation is induced in the solution zone due to the movement of the heater. Thus, the growth and the partial dissolution of the crystal occur periodically, depending upon the number of temperature oscillations set for a particular growth run during the process.

Table. 1. Temperature oscillation parameters and their impact on the number of grains in CdZnTe crystal

| Run | Heating rate (°C/hr) | Oscillation amp (°C) | Cooling rate (°C/hr) | Dwell time (hr) | No. of oscillation | No. of CdZnTe grains | |
|---|---|---|---|---|---|---|---|
| | | | | | | Bottom | Top |
| 1 | - | - | - | - | - | 14 | 8 |
| 2 | 2 | 3 | 0.67 | 18 | 5 | 16 | 2 |
| 3 | 20 | 10 | 1.05 | 14 | 10 | 5 | 2 |
| 4 | 15 | 15 | 1.36 | 12 | 40 | 1 | 1 |

After completion of crystal growth, the grown crystal was carefully removed from the quartz ampoule and thin Te top layer on the crystal surface was cleaned. Fig. 2a displays the photograph of the CdZnTe crystal grown in Run-4. The optical micrograph of the wafer cut across CdZnTe / Te interface is shown in Fig. 2b and it reveals a nearly flat growth interface with respect to CdZnTe crystal. Further, the Fig. 2c shows the microscopically flat interface observed through an IR transmission optical micrograph. The shape of the solid-liquid interface

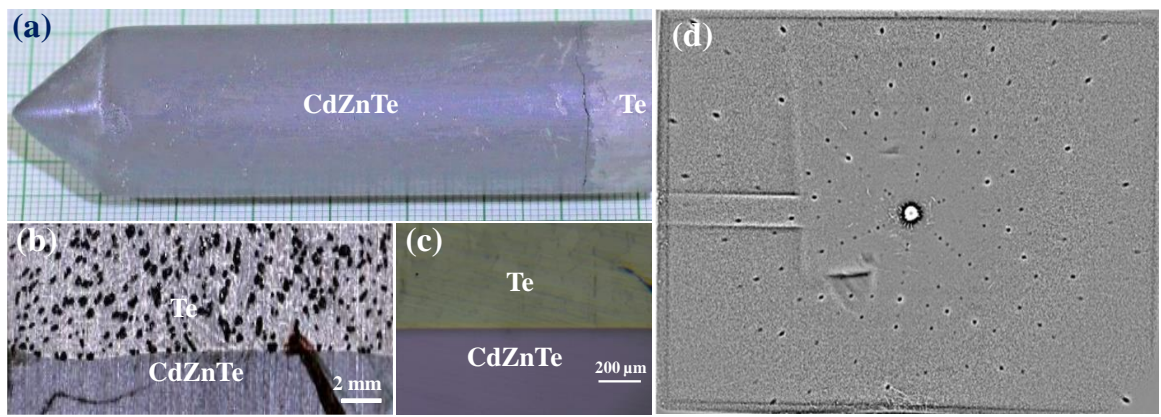

Fig.2. (a) Photograph of the CdZnTe crystal grown by Run-4. (b) The optical micrograph of the CdZnTe-Te interface at macroscopic level. (c) A magnified part of the growth interface as observed by IR microscope. (d) Back scattered Laue diffraction pattern of CdZnTe single crystal

is crucial in determining the quality of the crystal grown from the melt. The observed nearly flat interface suggests minimal defect formation in the crystal. Fig. 2d depicts the Laue diffraction pattern corresponding to the (111) orientation, exhibiting three-fold symmetry. The sharp diffraction pattern confirms the crystalline nature of the grown crystal.

After the growth process, the crystals were cut perpendicular to the growth axis at both the top and bottom sections using a diamond wire cutter to assess the impact of temperature fluctuations on the grain structure. Grain distribution analysis was conducted on these wafers using an optical microscope. After surface cleaning, the grain structures on the as-cut wafers were identified and counted, as shown in Fig. 3. Figs. 3(b–h) represent the images of the sliced wafers from both the bottom (just above the conical tapered region, as indicated in Fig. 3a) and the top of the crystal (approximately 10 mm below the Te–CdZnTe crystal interface) for four different growth runs. In the absence of temperature oscillation (Run-1), the bottom wafer (Fig. 3c) contained around 14 grains, whereas the top wafer had about 8 grains, with a larger grain at the wafer's center (Fig. 3b). For Run-2, which involved five temperature cycles with a $\Delta T$ of 3 °C, the bottom wafer contained 16 grains, including a large central grain (Fig. 3e), while the top wafer had only 2 grains (Fig. 3d). In Run-3, the bottom and top wafers contained approximately 6 and 1 grain respectively, as shown in Figs. 3g and 3f. In Run-4, a single-grain CdZnTe crystal was observed on both the bottom and top wafers, as shown in Figs. 3i and 3h, respectively.

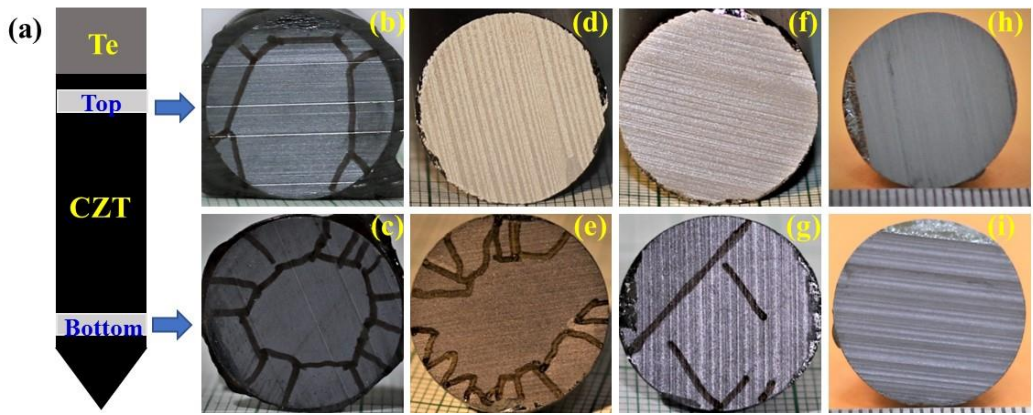

Fig.3. (a) A graphical representation of the grown crystal showing the positions of the bottom and top wafers. The cross-sectional optical micrographs of CdZnTe wafers sliced at bottom (c,e,g,i) and top (b,d,f,h) regions of the crystal from growth runs, (b,c) Run-1, (d,e) Run-2, (f,g) Run-3 and (h,i) Run-4

The role of temperature oscillation in the observed grain structures on these crystals can be understood as follows: In Run-1, spontaneous nucleation occurs at a small volume in the conically tapered region and it evolves as multi-grain structure. It should be noted here that multigrain structure is observed on both bottom and top portion of the crystal. However, in the middle portion, the wafer consists of a large grain with an additional tiny grain near periphery [24]. In Run-2, five cycles of temperature oscillation with a ΔT of 3 °C were applied at the beginning of the growth process over five days. During this period, the heater assembly was moved upward by ~ 20 mm (4 mm/day), allowing the CdZnTe crystal to grow by a similar length, slightly above the tapered region of the quartz ampoule. However, this minor temperature oscillation of 3 °C was insufficient to minimize the number of nucleation in the tapered region, resulting in a polycrystalline structure extending to the top wafer. To prevent the formation of a multi-grain structure near the tapered region, Run-3 involved increasing the oscillation amplitude to 10 °C and extending the cycles to 10. This approach helped reduce secondary nucleation and enhance grain size in the bottom portion of the crystal. As a result, the number of grains decreased to five at the bottom, with a single-grain structure forming at the top (Figs. 3g & 3f). Further increasing the number of temperature oscillations to 40 cycles, with an oscillation amplitude of 15 °C, facilitated the formation of a single grain throughout the entire length of the crystal. The sequential heating and cooling process promoted the formation of a single nucleation site. During heating, a few layers of the grown crystal dissolved, and upon cooling, they re-grew, effectively suppressing multi-nucleation during growth. After the successful growth of the crystal in Run-4, wafers from the bottom and top sections were analyzed to assess their structural integrity and compositional uniformity.

## 2.2. Characterization

Optical transmission spectroscopy measurements on polished wafers with a thickness of 2 mm were performed at room temperature using the Perkin Elmer Lambda 35 in the wavelength range of 700 -1100 nm. ALPHA-BRUKER spectrophotometer was utilized to record FTIR transmission spectra within the range of 500 - 6000 $cm^{-1}$. Photoluminescence (PL) spectra were recorded using Renishaw in-Via micro-Raman spectrometer having 1800 lines/mm grating with wavelength of 532 nm in the back scattering geometry. The sizes of Te inclusions and its statistical distribution in CdZnTe wafers were analyzed using IR transmission microscopy. A large field-of-view IR microscope objective, a CCD camera, an X-Y translation stage, and a 940 nm infrared light source with a wide-beam condenser were used for IR transmission microscopy setup. CdZnTe wafers were polished by mechanical and chemo-mechanical processes before electrode deposition in order to prepare for both electrical and gamma spectroscopy investigations.

The electroless deposition approach was used to coat Au layer on the surface-processed CdZnTe wafer. After five minutes of passivation with an $H_2O_2$ solution, the detector elements and electrodes are annealed for several hours at 100 °C [24–26]. A Keithley 6517 B electrometer was used to measure the electrical resistivity at room temperature in order to evaluate the electrical properties. Gamma spectroscopy measurements were carried out with the use of nuclear instrument module, which included a spectroscopic amplifier and a high voltage module. CdZnTe detector element with dimension of $10 \times 10 \times 5$ mm³ is coated with gold electrode on five sides to form a quasi-hemispherical detector. The top circular electrode with a diameter of 1 mm serves as anode while the remaining five faces act as cathode. The detector element was housed in a detector enclosure with 0.36 mm thick window made up of aluminium foil. The contacts to the anode and cathode were made using the silver epoxy. A USB-based multichannel analyzer was used for spectrum collection, acquiring data for 300 seconds at a bias of 1000 V and a shaping time of 1 μs. The spectroscopic performance was evaluated using gamma radiation sources, specifically $^{137}$Cs and $^{133}$Ba.

3. Results

3.1. NIR and mid-IR transmission spectroscopy

NIR transmission spectra of the CdZnTe wafers taken from bottom and top portion of the crystal are shown in Fig. 4a. This result indicates that the cut-off wavelength remains constant about 815 nm while the transmission percentage varies slightly for wafers from top and bottom regions of the crystal. Fig. 4b shows the mid-IR transmission spectra of the wafers in the wavelength range of 5 - 28 μm. Moreover, the high optical transmission of ~ 60 % in mid-IR wavelength range indicates the good structural quality of these wafers.

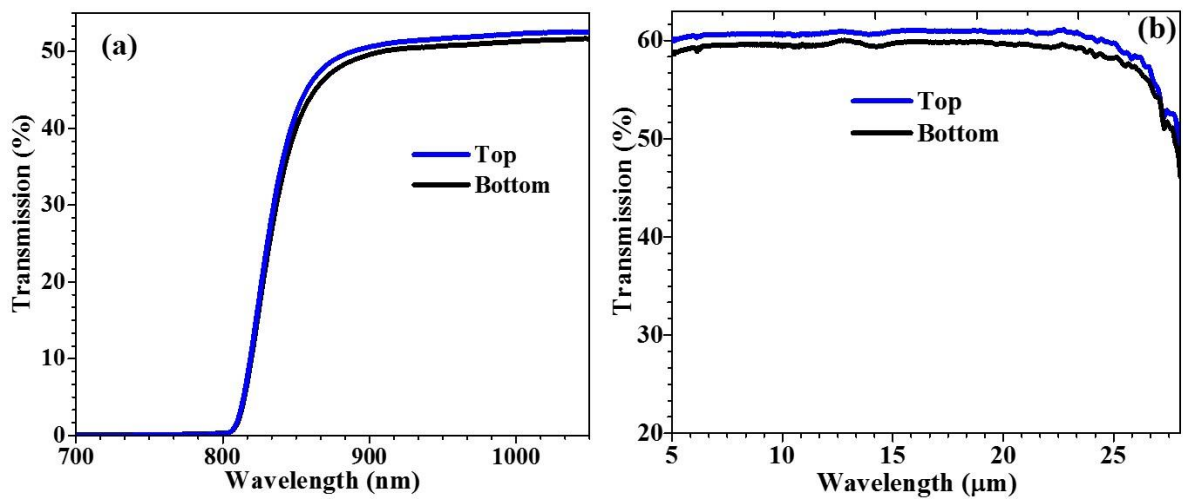

Fig 4. (a) Near IR and (b) mid-IR transmission spectra of CdZnTe wafers from bottom and top portion of the crystal.

## 3.2. Photoluminescence spectroscopy

The room temperature PL spectra recorded on the bottom and top regions of the CdZnTe crystal are shown in Fig. 5. Both wafers exhibit a broad single emission peak with asymmetry on the higher energy side. The wavelength of PL peak is found to be 790.9 and 790.3 nm for these wafers. The PL peak arises mainly due to free exciton whose binding energy is about 10.3 meV lower than the band gap of $Cd_{1-x}Zn_xTe$ with x < 0.10 [27]. Thus, the bandgap of the material is calculated by adding this exciton binding energy to the measured peak value. The bandgap of these wafers is found to be ~ 1.58 eV which is very close to the theoretical value of 1.59 eV [28]. Further, the nearly constant PL peak position also confirms the composition uniformity at top and bottom of the crystal.

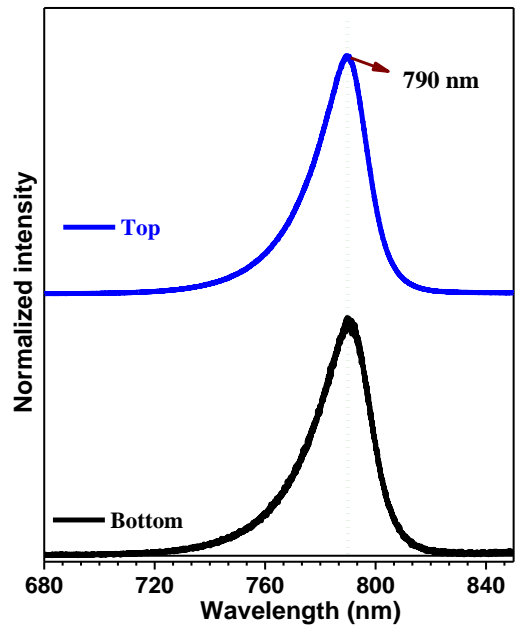

Fig. 5. Photoluminescence spectra of CdZnTe wafers from bottom and top portion of the crystal

## 3.3. Infra-red microscopy analysis

Figs. 6a and 6b depict the IR transmission microscopic images of CdZnTe wafers taken at bottom and top of the crystal, respectively. The dark spots in these images are formed by Te inclusions which reflect the IR source while the matrix transmit IR source. The Figs. 6c and 6d displays the statistical size distribution of Te inclusion in these wafers. Here, a nearly uniform distribution of Te inclusion is observed for the wafers from bottom and top of CdZnTe wafers with an average diameter of 9.2 ± 0.3 and 9.8 ± 0.1 μm, respectively. The Te inclusion density is found to be 1.34 and 1.38 × $10^4$ /cm$^2$ for the wafers from bottom and top portion of the crystal, respectively. The lateral size of Te inclusions is less than 16 μm. The observed Te distribution is comparable to the reports available in literature for detector grade CdZnTe crystal [29]. Additionally, Te inclusion mapping along the growth axis of the crystal from Run-3 was performed using IR microscopy at four distinct positions: bottom, lower-middle (1), upper-middle (2), and top regions, as shown in Figs. 7(a)–(d). The corresponding IR micrographs and statistical size distributions of Te inclusions at these locations are presented in Figs. 7(e)–(h). The Te inclusion density is found to be 1.12 × $10^5$, 5.02 × $10^4$, 4.75 × $10^4$, and 4.20 × $10^4$ /cm$^2$ for the wafers from bottom, lower-middle (1), upper-middle (2), and top regions, respectively. IR microscopy reveals a slightly higher concentration of Te inclusions in the bottom wafer, with a gradual decrease observed toward the top end of the crystal.

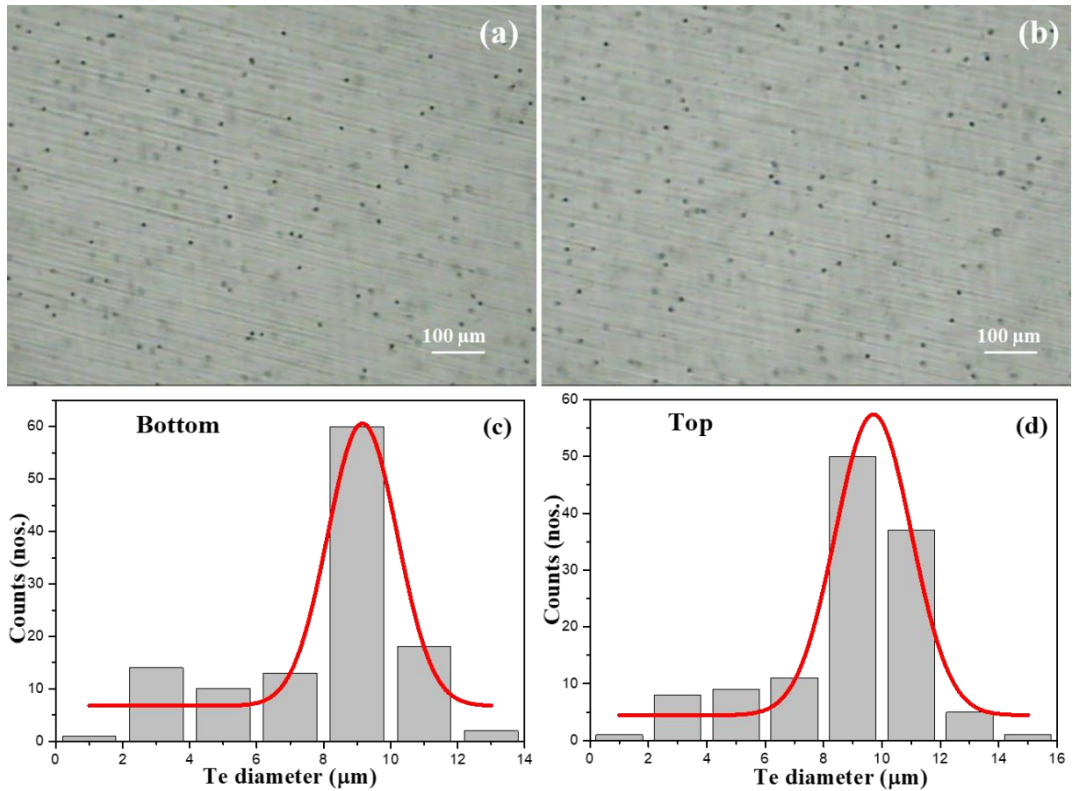

Fig. 6. IR transmission microscopic image of the CdZnTe crystal at (a) bottom and (b) top regions of the crystal. (c-d) Statistical size distribution of *Te* inclusion extracted from the images given in 6 (a-b), respectively.

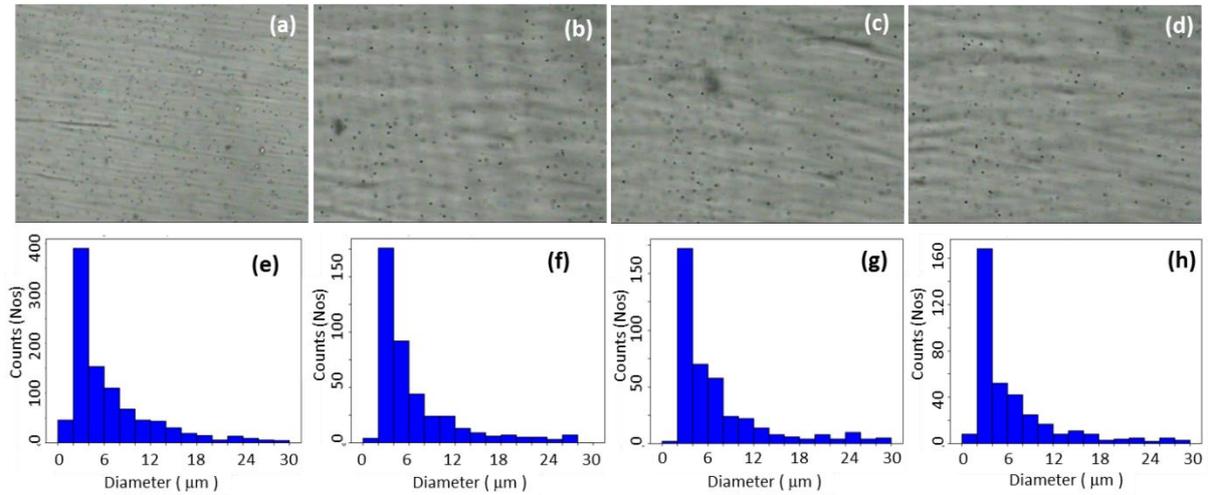

Fig.7. Tellurium inclusions images taken at different position in the grown CZT crystal (a) bottom, (b) lower middle 1, (c) upper middle 2 and (d) Top. Their respective distributions are shown in (e)-(h).

### 3.4. I-V characteristics

The resistivity of these CdZnTe wafers were assessed using I-V characteristics which is shown in Fig. 8. The resistivity of the CdZnTe was found to be $1.5 \times 10^9$, and $2.2 \times 10^9$ $\Omega \cdot cm$ for the wafers from bottom and top region of the crystal, respectively. A small deviation in the resistivity among these wafers emphasizes the nearly homogeneous distribution of defects in the CdZnTe crystal.

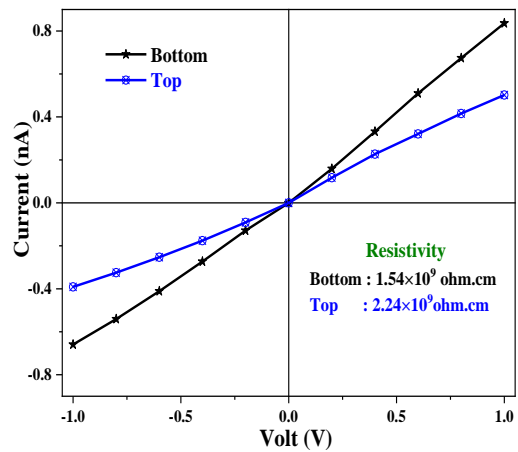

Fig.8. I-V characteristics of CdZnTe crystals at different regions of the grown crystal

## 3.5. Gamma spectrum analysis

Fig. 9a shows the detector response for the $^{137}$Cs gamma radiation source. The detector energy resolution is found to be ~ 4.5 % at the 662 keV photo peak of $^{137}$Cs using quasi-hemispherical detector geometry. Also, the photo peak was measured at different bias with an energy of 81 keV using $^{133}$Ba source from the detector in planner geometry and the spectrum is shown in Fig. 9b. The mobility-lifetime ($\mu\tau$) product is estimated using the Hecht equation that gives the total charge collected in a negatively biased condition [30].

$$Q(V) = [N_0 q V (\mu\tau)_e / d^2] \times [1 - \exp(d^2/(\mu\tau)_e V)] \quad (1)$$

Here, $N_0$ is the number of charge carriers created by the incident gamma photon, $Q$ - the total charge collected, $d$ - thickness of the detector, $q$ - electronic charge, and $V$ - applied bias. Using

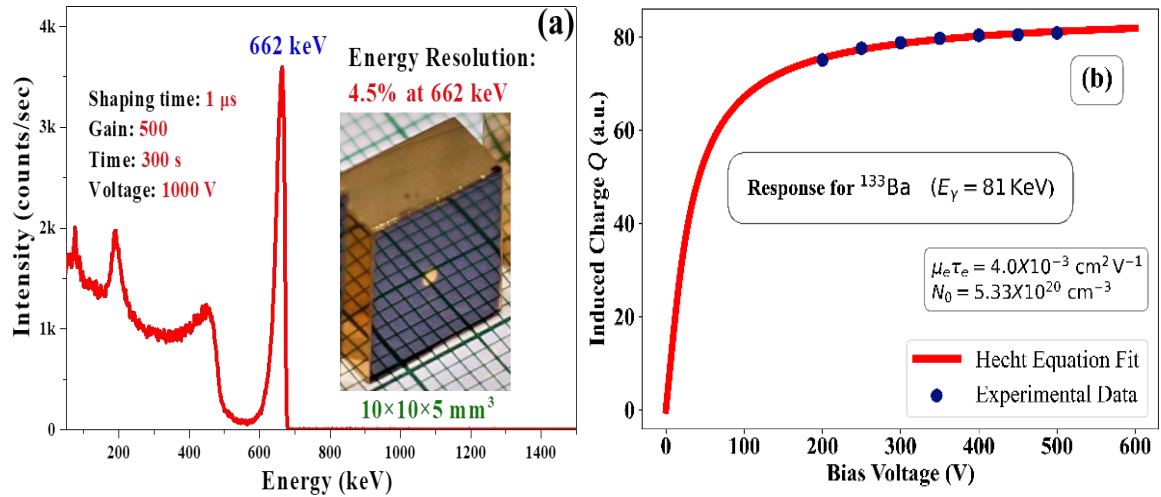

Fig. 9 (a) The CdZnTe detector response to the $^{137}$Cs gamma radiation, and (b) induced charge as a function of applied bias by CdZnTe planar detector for 81 keV $^{133}$Ba source.

the best fit parameters from Equation 1, the $\mu\tau$ product is estimated to be ~ 4.0×10$^{-3}$ cm²/V, which is comparable to values reported in the literature for CdZnTe crystal [28,31].

## 4. Discussion

Based on the above results, we confirm that the CdZnTe crystal has been grown with reasonably high quality. FTIR and PL spectroscopy studies validate the compositional uniformity at bottom and top regions of the crystal. Additionally, the distribution of Te inclusions is nearly uniform at bottom and top regions of the crystal. Furthermore, the uniformity in composition and defects is evident from the nearly constant resistivity observed in the CdZnTe wafers. The gamma energy resolution of the CdZnTe crystal (4.5% at 662 keV) grown with temperature oscillation in THM is consistent with that of CdZnTe crystals grown without using temperature oscillation during THM growth [24].

The primary objective of this study was to grow large-grain CdZnTe single crystals, enabling the extraction of multiple detector elements even from a small-diameter crystal. Optical microscopy studies clearly show that applying temperature oscillation during growth increases grain size while reducing the number of grains. During the ramp-up phase of the thermal cycle, the already crystallized CdZnTe undergoes partial dissolution, while during the ramp-down phase, the solution becomes slightly undersaturated. Towards the end of the ramp-down and during the isothermal phase of thermal cycling, supersaturation occurs near the growth interface due to the continuous upward movement of the heater and subsequently, growth of CdZnTe takes place. Because of the low thermal conductivity of both the solution and the grown CdZnTe crystal, there is a high likelihood of local superheating / supercooling near the growth interface. This condition can lead to multi-nucleation at any point during the crystal growth process. However, the secondary nucleated CdZnTe grains dissolve during thermal cycling, ultimately fostering the formation of large-grain CdZnTe single crystals. Thus, we attribute the growth of large-grain single crystals to the periodic temperature oscillation applied during the crystal growth process.

## 5. Conclusion

Large grain CdZnTe single crystal was successfully grown using temperature oscillation method under conventional travelling heater method (THM). The temperature oscillation above the growth temperature effectively controls nucleation at the crucible bottom region by dissolving the multi-nuclei formation and also, it removes any secondary nucleation along the length of the crystal during growth. The uniformity of structure and composition were confirmed by the observation of near uniform bandgap of 1.58 eV, electrical resistivity of $\sim 10^9$ $\Omega$.cm, optical transmittance and Te inclusion density at bottom and top regions of the grown crystal. The energy resolution of the radiation detector was determined to be 4.5 % at 662 keV using a unique quasi-hemispherical electrode. The mobility-lifetime product was found to be $4.0 \times 10^{-3}$ cm$^2$/V with an energy of 81 keV using $^{133}$Ba source in planner geometry. The results of this research highlight the significance of utilizing temperature oscillation in a self-nucleating approach in THM for producing detector-grade CdZnTe single crystals.

## 6. Acknowledgement

The authors thank Dr. Vidhya Sundararajan, Associate Director, SQRMG / IGCAR and Shri C.G. Karhadkar, Director, IGCAR for their support and encouragement. One of the authors, P. Vijayakumar, is thankful to IGCAR for the award of Research Associate and Visiting Scientist fellowships.

## 7. References

[1] A.E. Bolotnikov, G.S. Camarda, E. Chen, S. Cheng, Y. Cui, R. Gul, R. Gallagher, V. Dedic, G. De Geronimo, L. Ocampo Giraldo, J. Fried, A. Hossain, J.M. Mackenzie, P. Sellin, S. Taherion, E. Vernon, G. Yang, U. El-Hanany, R.B. James, CdZnTe position-sensitive drift detectors with thicknesses up to 5 cm, Appl. Phys. Lett. 108 (2016) 4–8. doi:10.1063/1.4943161.

[2] M.D. Alam, S.S. Nasim, S. Hasan, Recent progress in CdZnTe based room temperature detectors for nuclear radiation monitoring, Prog. Nucl. Energy. 140 (2021) 103918. doi:10.1016/j.pnucene.2021.103918.

[3] A.E. Bolotnikov, G.S. Camarda, G. De Geronimo, J. Fried, D. Hodges, A. Hossain, K. Kim, G. Mahler, L.O. Giraldo, E. Vernon, G. Yang, R.B. James, A 4×4 array module of position-sensitive virtual Frisch-grid CdZnTe detectors for gamma-ray imaging spectrometers, Nucl. Instruments Methods Phys. Res. Sect. A Accel. Spectrometers, Detect. Assoc. Equip. 954 (2020) 161036. doi:10.1016/j.nima.2018.07.090.

[4] I. Kuvvetli, C. Budtz-Jørgensen, E. Caroli, N. Auricchio, CZT drift strip detectors for high energy astrophysics, Nucl. Instruments Methods Phys. Res. Sect. A Accel. Spectrometers, Detect. Assoc. Equip. 624 (2010) 486–491. doi:10.1016/j.nima.2010.03.172.

[5] F.P. Doty, J.F. Butler, J.F. Schetzina, K.A. Bowers, Properties of CdZnTe crystals grown by a high pressure Bridgman method, J. Vac. Sci. Technol. B Microelectron. Nanom. Struct. Process. Meas. Phenom. 10 (1992) 1418–1422. doi:10.1116/1.586264.

[6] T. Asahi, O. Oda, Y. Taniguchi, A. Koyama, Growth and characterization of 100 mm diameter CdZnTe single crystals by the vertical gradient freezing method, J. Cryst. Growth. 161 (1996) 20–27. doi:10.1016/0022-0248(95)00606-0.

[7] K. Strzałkowski, The composition effect on the thermal and optical properties across CdZnTe crystals, J. Phys. D. Appl. Phys. 49 (2016). doi:10.1088/0022-3727/49/43/435106.

[8] B. Hong, S. Zhang, L. Zheng, H. Zhang, C. Wang, B. Zhao, Studies on thermal and interface optimization for CdZnTe crystals by unseeded Traveling Heater Method, J. Cryst. Growth. 546 (2020). doi:10.1016/j.jcrysgro.2020.125776.

[9] M. Ünal, · Özden Başar Balbaşı, · Mehmet, C. Karaman, · Ayşe, M. Genç, · Mehmet Parlak, R. Turan, Production of High-Performance CdZnTe Crystals Grown by THM for Radiation Detection Applications, J. Electron. Mater. 51 (2022) 4675–4680. doi:10.1007/s11664-022-09663-y.

[10] B. Hong, S. Zhang, L. Zheng, H. Zhang, C. Wang, B. Zhao, Controlling nucleation during unseeded THM growth of CdZnTe crystal, J. Cryst. Growth. 534 (2020). doi:10.1016/j.jcrysgro.2020.125482.

[11] N. Zhang, A. Yeckel, J.J. Derby, Maintaining convex interface shapes during electrodynamic gradient freeze growth of cadmium zinc telluride using a dynamic, bell-curve furnace profile, J. Cryst. Growth. 355 (2012) 113–121. doi:10.1016/j.jcrysgro.2012.06.042.

[12] E. Saucedo, P. Rudolph, E. Dieguez, Modified Bridgman growth of CdTe crystals, J. Cryst. Growth. 310 (2008) 2067–2071. doi:10.1016/j.jcrysgro.2007.11.181.

[13] V. Carcelén, N. Vijayan, E. Diéguez, A. Zappettini, M. Zha, L. Sylla, A. Fauler, M. Fiederle, New approaches in order to enlarge the grain size of bulk CdZnTe (CZT) crystals, J. Optoelectron. Adv. Mater. 10 (2008) 3135–3140.

[14] V. Carcelén, K.H. Kim, G.S. Camarda, A.E. Bolotnikov, A. Hossain, G. Yang, J. Crocco, H. Bensalah, F. Dierre, E. Diéguez, R.B. James, Pt coldfinger improves quality of Bridgman-grown Cd 0.9Zn 0.1Te:Bi crystals, J. Cryst. Growth. 338 (2012) 1–5. doi:10.1016/j.jcrysgro.2011.09.031.

[15] M. Schieber, W.F. Schnepple, L. Van den Berg, Vapor growth of HgI2 by periodic source or crystal temperature oscillation, J. Cryst. Growth. 33 (1976) 125–135. doi:10.1016/0022-0248(76)90087-7.

[16] F.A.S. Al-Alamy, A.A. Balchin, Applications of the temperature oscillation method to the growth of layer compounds by iodine vapour transport, J. Cryst. Growth. 39 (1977) 275–286. doi:10.1016/0022-0248(77)90275-5.

[17] A.M.J.G. Van Run, Computation of striated impurity distributions in melt-grown crystals, taking account of periodic remelt, J. Cryst. Growth. 47 (1979) 680–692. doi:10.1016/0022-0248(79)90012-5.

[18] A.M.J.G. Van Run, A critical pulling rate for remelt suppression in silicon crystal growth, J. Cryst. Growth. 53 (1981) 441–442. doi:10.1016/0022-0248(81)90097-X.

[19] M.J. Crochet, F.T. Geyling, J.J. Van Schaftingen, Numerical simulation of the horizontal Bridgman growth of a gallium arsenide crystal, J. Cryst. Growth. 65 (1983) 166–172. doi:10.1016/0022-0248(83)90049-0.

[20] V.M. Masalov, G.A. Emel'chenko, A.B. Mikhajlov, Hydrodynamics and oscillation of temperature in single crystal growth from high-temperature solutions with use of ACRT, J. Cryst. Growth. 119 (1992) 297–302. doi:10.1016/0022-0248(92)90682-9.

[21] J.R. Carruthers, Origins of convective temperature oscillations in crystal growth melts, J. Cryst. Growth. 32 (1976) 13–26. doi:10.1016/0022-0248(76)90004-X.

[22] R.S. Mroczkowski, A.F. Witt, H.C. Gatos, Effects of Back-Melting on the Dislocation Density in Single Crystals: GaSb, J. Electrochem. Soc. 115 (1968) 545. doi:10.1149/1.2411323.

[23] G. Attolini, P. Ferro, G. Trevisi, Growth of germanium crystals by time-varying temperature profile, CVT methods, Mater. Sci. Eng. B. 286 (2022) 116045. doi:https://doi.org/10.1016/j.mseb.2022.116045.

[24] P. Vijayakumar, E.P. Amaladass, K. Ganesan, R.M. Sarguna, V. Roy, S. Ganesamoorthy, Development of travelling heater method for growth of detector grade CdZnTe single crystals, Mater. Sci. Semicond. Process. 169 (2024) 107897. doi:10.1016/j.mssp.2023.107897.

[25] G. Yang, W. Jie, Q. Li, Effects of annealing on the properties of Au-Cd0.9Zn 0.1Te contacts, Mater. Sci. Eng. B Solid-State Mater. Adv. Technol. 123 (2005) 172–175.

doi:10.1016/j.mseb.2005.07.024.

[26] P. Vijayakumar, E.P. Amaladass, K. Ganesan, R.M. Sarguna, S. Chinnathambi, S. Ganesamoorthy, V. Sridharan, A. Mani, N. Subramanian, Electrical resistivity studies on Cd0.9Zn0.1Te single crystals grown by travelling heater method, AIP Conf. Proc. 2265 (2020) 9–12. doi:10.1063/5.0018571.

[27] K.S. and J.L.P. J.M. Francou, Shallow donors in CdTe, Phys. Rev. B. 41 (1990) 35–46.

[28] M. Ünal, Ö.B. Balbaşı, S.H. Sedani, M.C. Karaman, G. Çelik, D. Bender, A.M. Genç, M. Parlak, R. Turan, Production of detector grade CdZnTe crystal with VGF furnace by analyzing segregation of Zn and In, J. Cryst. Growth. 615 (2023). doi:10.1016/j.jcrysgro.2023.127236.

[29] B. Ren, J. Zhang, M. Liao, J. Huang, L. Sang, Y. Koide, L. Wang, High-performance visible to near-infrared photodetectors by using (Cd,Zn)Te single crystal, Opt. Express. 27 (2019) 8935. doi:10.1364/oe.27.008935.

[30] U.N. Roy, A. Gueorguiev, S. Weiller, J. Stein, Growth of spectroscopic grade Cd0.9Zn0.1Te:In by THM technique, J. Cryst. Growth. 312 (2009) 33–36. doi:10.1016/j.jcrysgro.2009.09.035.

[31] Y.D. Xu, W.Q. Jie, Y.H. He, R.R. Guo, T. Wang, G.Q. Zha, Size and distribution of Te inclusions in detector-grade CdZnTe ingots, Prog. Nat. Sci. Mater. Int. 21 (2011) 66–72. doi:10.1016/S1002-0071(12)60027-6.